\documentstyle[epsf]{article}

\newcommand{\be}{\begin{equation}}
\newcommand{\ee}{\end{equation}}
\newcommand{\bea}{\begin{eqnarray}}
\newcommand{\eea}{\end{eqnarray}}
\newcommand{\bi}{\bibitem}
\newcommand{\nn}{\nonumber}

\newcommand{\complex}{{{\rm I} \kern -.59em {\rm C}}}

\makeatletter
\let\chapter\hid@chapter
\makeatother

\begin{document}
\begin{titlepage}
  \renewcommand{\thefootnote}{\fnsymbol{footnote}}
  \begin{flushright}
    \begin{tabular}{l@{}}
      IFUNAM-FT98-1\\
      hep-ph/9802280
    \end{tabular}
  \end{flushright}

  \vskip 0pt plus 0.4fill

  \begin{center}
    {\LARGE \textbf{Constraints on Finite Soft Supersymmetry
  Breaking Terms}\footnote{Supported partly by the mexican projects
    CONACYT 3275-PE, PAPIIT IN110296, by the EU projects
    FMBI-CT96-1212 and ERBFMRXCT960090, and by the Greek project
    PENED95/1170;1981.}\par}

  \end{center}

  \vskip 0pt plus 0.2fill

  \begin{center}
    {\large
      T.~Kobayashi%
      }\\
    \textit{
      Inst.~of Particle and Nuclear Studies,
      Tanashi, Tokyo 188, Japan
      }\\
    \vspace{1ex}
    {\large 
      J.~Kubo%
      }\\
    \textit{Dept.~of Physics, Kanazawa Univ.~,
      Kanazawa 920-1192, Japan
      }\\
    \vspace{1ex}
    {\large
      M.~Mondrag\'on%
      }\\
    \textit{Inst.~de F{\'{\i}}sica, UNAM, Apdo. Postal 20-364,
      M\'exico 01000 D.F., M\'exico
      }\\
    \vspace{1ex}
    {\large
      G.~Zoupanos%
        }\\
    \textit{
      Physics Dept., Nat. Technical Univ.,
      GR-157 80 Zografou, Athens, Greece\\
    and\\
      Institut f.~Physik, Humboldt-Universit\"at, D10115 Berlin,
      Germany
      }
   
    \vskip 1ex plus 0.3fill

    {\large February 6, 1998}

    \vskip 1ex plus 0.7fill

    \textbf{Abstract}
  \end{center}
  \begin{quotation}
  A new solution to the requirement of two-loop finiteness of the soft
  supersymmetry breaking terms (SSB) parameters is found in
  Finite-Gauge-Yukawa unified theories. The new solution has the form
  of a sum rule for the relevant scalar masses, relaxing the
  universality required by the previously known solution, which leads
  to models with unpleasant phenomenological consequences. Using the
  sum rule we investigate two Finite-Gauge-Yukawa unified models and
  we determine their spectrum in terms of few parameters. Some
  characteristic features of the models are that a) the old agreement
  of the top quark mass prediction with the measured value remains
  unchanged, b) the lightest Higgs mass is predicted around 120 GeV,
  c) the s-spectrum starts above 200 GeV.
    \end{quotation}

  \vskip 0pt plus 2fill

  \setcounter{footnote}{0}

\end{titlepage}

\section{Introduction}

Finite Unified Theories (FUTs) have fundamental conceptual importance
in the search of the final theory describing Elementary Particle
Physics. 

In our attempts here, following the usual minimality assumption, we
are restricted in unifying only the known gauge interactions, and it
is interesting to point out that {\em finiteness does not require
  gravity}.  Finiteness is based on the fact that it is possible to
find renormalization group invariant (RGI) relations among couplings
that keep finiteness in perturbation theory, even to all orders
\cite{acta}.

\section{Finiteness and Reduction of Couplings in $N=1$ SUSY Gauge Theories}
 
Let us then consider a chiral, anomaly free,
$N=1$ globally supersymmetric
gauge theory based on a group G with gauge coupling
constant $g$. The
superpotential of the theory is given by
\bea
W&=& \frac{1}{2}\,m^{ij} \,\Phi_{i}\,\Phi_{j}+
\frac{1}{6}\,C^{ijk} \,\Phi_{i}\,\Phi_{j}\,\Phi_{k}~,
\label{supot}
\eea
where $m^{ij}$ and $C^{ijk}$ are gauge invariant tensors and
the matter field $\Phi_{i}$ transforms
according to the irreducible representation  $R_{i}$
of the gauge group $G$. 

 The one-loop $\beta$-function of the gauge
coupling $g$ is given by 
\bea
\beta^{(1)}_{g}&=&\frac{d g}{d t} =
\frac{g^3}{16\pi^2}\,[\,\sum_{i}\,l(R_{i})-3\,C_{2}(G)\,]~,
\label{betag}
\eea
where $l(R_{i})$ is the Dynkin index of $R_{i}$ and $C_{2}(G)$
 is the
quadratic Casimir of the adjoint representation of the
gauge group $G$. The $\beta$-functions of
$C^{ijk}$,
by virtue of the non-renormalization theorem, are related to the
anomalous dimension matrix $\gamma^j_i$ of the matter fields
$\Phi_{i}$ as:
\be
\beta_{C}^{ijk}=\frac{d}{dt}\,C^{ijk}=C^{ijp}\,
~\sum_{n=1}~\frac{1}{(16\pi^2)^n}\,\gamma_{p}^{k(n)} +(k
\leftrightarrow i) +(k\leftrightarrow j)~.
\label{betay}
\ee
At one-loop level $\gamma^j_i$ is 
\be
\gamma_i^{j(1)}=\frac{1}{2}C_{ipq}\,C^{jpq}-2\,g^2\,C_{2}(R_{i})\delta_i^j~,
\label{gamay}
\ee
where $C_{2}(R_{i})$ is the quadratic Casimir of the representation
$R_{i}$, and $C^{ijk}=C_{ijk}^{*}$.

As one can see from Eqs.~(\ref{betag}) and (\ref{gamay}) 
 all the one-loop $\beta$-functions of the theory vanish if
 $\beta_g^{(1)}$ and $\gamma_i^{j(1)}$ vanish, i.e.
\be
\sum _i \ell (R_i) = 3 C_2(G) \,,~~~~~~~~
\frac{1}{2}C_{ipq} C^{jpq} = 2\delta _i^j g^2  C_2(R_i)\,.
\label{2nd}
\ee

A very interesting result is that the conditions (\ref{2nd}) are
necessary and sufficient for finiteness at
the two-loop level.

The one- and two-loop finiteness conditions (\ref{2nd}) restrict
considerably the possible choices of the irreps.~$R_i$ for a given
group $G$ as well as the Yukawa couplings in the superpotential
(\ref{supot}).  Note in particular that the finiteness conditions cannot be
applied to the supersymmetric standard model (SSM), since the presence
of a $U(1)$ gauge group is incompatible with the condition
(\ref{2nd}), due to $C_2[U(1)]=0$.  This naturally leads to the
expectation that finiteness should be attained at the grand unified
level only, the SSM being just the corresponding, low-energy,
effective theory.

A natural question to ask is what happens at higher loop orders.  The
finiteness conditions (\ref{2nd}) impose relations between gauge and
Yukawa couplings.  We would like to guarantee that such relations
leading to a reduction of the couplings hold at any renormalization
point.  The necessary, but also sufficient, condition
for this to happen is to require that such relations are solutions to
the reduction equations (REs) 
\be \beta_g {d C^{ijk}\over dg} = \beta^{ijk}
\label{redeq2}
\ee
and hold at all orders.  Remarkably the existence of
all-order power series solutions to (\ref{redeq2}) can be decided at
the one-loop level.

There exists an all-order finiteness theorem \cite{LPS} 
based on: (a) the structure of the supercurrent in $N=1$ SYM and on 
(b) the non-renormalization properties of $N=1$ chiral anomalies
\cite{LPS}. 
Details on the proof can be found in ref.~\cite{LPS}.

One-loop finiteness implies that the Yukawa couplings $C^{ijk}$
must be functions of the gauge coupling $g$. To find a similar
condition to all orders it is necessary and sufficient for the Yukawa
couplings to be a formal power series in $g$, which is solution of the
REs (\ref{redeq2}).

\section{One and two-loop finite supersymmetry breaking terms}

Here we would like to use the two-loop RG functionsm to re-investigate
their two-loop finiteness and derive
the two-loop soft scalar-mass sum rule.

Consider the superpotential given by (\ref{supot}) 
along with the Lagrangian for SSB terms,
\bea
-{\cal L}_{\rm SB} &=&
\frac{1}{6} \,h^{ijk}\,\phi_i \phi_j \phi_k
+
\frac{1}{2} \,b^{ij}\,\phi_i \phi_j
+
\frac{1}{2} \,(m^2)^{j}_{i}\,\phi^{*\,i} \phi_j+
\frac{1}{2} \,M\,\lambda \lambda+\mbox{H.c.}~
\eea
where the $\phi_i$ are the
scalar parts of the chiral superfields $\Phi_i$ , $\lambda$ are the gauginos
and $M$ their unified mass.
Since we would like to consider
only finite theories here, we assume that 
the gauge group is  a simple group and the one-loop
$\beta$ function of the 
gauge coupling $g$  vanishes.
We also assume that the reduction equations (\ref{redeq2}) 
admit power series solutions of the form
\bea 
C^{ijk} &=& g\,\sum_{n=0}\,\rho^{ijk}_{(n)} g^{2n}~,
\label{Yg}
\eea 
According to the finiteness theorem
of ref.~\cite{LPS}, the theory is then finite to all orders in
perturbation theory, if, among others, the one-loop anomalous dimensions
$\gamma_{i}^{j(1)}$ vanish.  The one- and two-loop finiteness for
$h^{ijk}$ can be achieved by 
\bea h^{ijk} &=& -M C^{ijk}+\dots =-M
\rho^{ijk}_{(0)}\,g+O(g^5)~.
\label{hY}
\eea

Now, to obtain the two-loop sum rule for 
soft scalar masses, we assume that 
the lowest order coefficients $\rho^{ijk}_{(0)}$ 
and also $(m^2)^{i}_{j}$ satisfy the diagonality relations
\bea
\rho_{ipq(0)}\rho^{jpq}_{(0)} &\propto & \delta_{i}^{j}~\mbox{for all} 
~p ~\mbox{and}~q~~\mbox{and}~~
(m^2)^{i}_{j}= m^{2}_{j}\delta^{i}_{j}~,
\label{cond1}
\eea
respectively.
Then we find the following soft scalar-mass sum
rule
\bea
(~m_{i}^{2}+m_{j}^{2}+m_{k}^{2}~)/
M M^{\dag} &=&
1+\frac{g^2}{16 \pi^2}\,\Delta^{(1)}+O(g^4)~
\label{sumr} 
\eea
for i, j, k with $\rho^{ijk}_{(0)} \neq 0$, where $\Delta^{(1)}$ is
the two-loop correction, which vanishes for the
universal choice in accordance with the previous findings of
ref.~\cite{jack3}.
 
\section{Finite Unified Theories}

A predictive Gauge-Yukawa unified SU(5) model which is finite to all
orders, in addition to the requirements mentioned already, should also
have the following properties:

\begin{enumerate}

\item 
One-loop anomalous dimensions are diagonal,
i.e.,  $\gamma_{i}^{(1)\,j} \propto \delta^{j}_{i} $,
according to the assumption (\ref{cond1}).

\item
Three fermion generations, $\overline{\bf 5}_{i}~~
(i=1,2,3)$, obviously should not couple to ${\bf 24}$.
This can be achieved for instance by imposing $B-L$ 
conservation.

\item
The two Higgs doublets of the MSSM should mostly be made out of a 
pair of Higgs quintet and anti-quintet, which couple to the third
generation.
\end{enumerate}
In the following we discuss two versions of the all-order finite
model.

\vspace{0.2cm}
\noindent
${\bf A}$:  The model of ref. \cite{kmz1}.
\newline
${\bf B}$:  A slight variation of  the 
 model ${\bf A}$.

The  superpotential which describe the two models 
takes the form \cite{kmz1,kkmz-npb}
\bea
W &=& \sum_{i=1}^{3}\,[~\frac{1}{2}g_{i}^{u}
\,{\bf 10}_i{\bf 10}_i H_{i}+
g_{i}^{d}\,{\bf 10}_i \overline{\bf 5}_{i}\,
\overline{H}_{i}~] +
g_{23}^{u}\,{\bf 10}_2{\bf 10}_3 H_{4} \\
 & &+g_{23}^{d}\,{\bf 10}_2 \overline{\bf 5}_{3}\,
\overline{H}_{4}+
g_{32}^{d}\,{\bf 10}_3 \overline{\bf 5}_{2}\,
\overline{H}_{4}+
\sum_{a=1}^{4}g_{a}^{f}\,H_{a}\, 
{\bf 24}\,\overline{H}_{a}+
\frac{g^{\lambda}}{3}\,({\bf 24})^3~,\nn
\label{super}
\eea
where 
$H_{a}$ and $\overline{H}_{a}~~(a=1,\dots,4)$
stand for the Higgs quintets and anti-quintets.

The non-degenerate and isolated solutions to $\gamma^{(1)}_{i}=0$ for
the models $\{ {\bf A}~,~{\bf B} \}$ are: 
\bea (g_{1}^{u})^2
&=&\{\frac{8}{5},\frac{8}{5} \}g^2~, ~(g_{1}^{d})^2
=\{\frac{6}{5},\frac{6}{5}\}g^2~,~
(g_{2}^{u})^2=(g_{3}^{u})^2=\{\frac{8}{5},\frac{4}{5}\}g^2~,\label{SOL5}\\
(g_{2}^{d})^2 &=&(g_{3}^{d})^2=\{\frac{6}{5},\frac{3}{5}\}g^2~,~
(g_{23}^{u})^2 =\{0,\frac{4}{5}\}g^2~,~
(g_{23}^{d})^2=(g_{32}^{d})^2=\{0,\frac{3}{5}\}g^2~,
\nn\\
(g^{\lambda})^2 &=&\frac{15}{7}g^2~,~ (g_{2}^{f})^2
=(g_{3}^{f})^2=\{0,\frac{1}{2}\}g^2~,~ (g_{1}^{f})^2=0~,~
(g_{4}^{f})^2=\{1,0\}g^2~.\nn 
\eea 
According to the theorem of
ref.~\cite{LPS} these models are finite to all orders.  After the
reduction of couplings the symmetry of $W$ is enhanced
\cite{kkmz-npb}.

The main difference of the models
${\bf A}$ and ${\bf B}$ is
that three pairs of Higgs quintets and anti-quintets couple to 
the ${\bf 24}$ for ${\bf B}$ so that it is not necessary 
to mix
them with $H_{4}$ and $\overline{H}_{4}$ in order to
achieve the triplet-doublet splitting after the symmetry breaking 
of $SU(5)$.

\section{Predictions of Low Energy Parameters}
 
Since the gauge symmetry is spontaneously broken
below $M_{\rm GUT}$, the finiteness conditions 
do not restrict the renormalization property at low energies, and
all it remains are boundary conditions on the
gauge and Yukawa couplings (\ref{SOL5})
and the $h=-MC$ relation (\ref{hY}) and
the soft scalar-mass sum rule (\ref{sumr}) at $M_{\rm GUT}$.
So we examine the evolution of these parameters according
to their renormalization group equations at two-loop 
for dimensionless parameters and 
 at one-loop 
for dimensional ones with
these boundary conditions.
Below $M_{\rm GUT}$ their evolution is assumed to be
governed by the MSSM. We further assume a unique 
supersymmetry breaking scale
$M_{s}$ so that
below $M_{s}$ the SM is the correct effective theory.

The predictions for the top quark mass $M_t$ are $\sim 183$ and $\sim
174$ GeV in models $\bf A$ and $\bf B$ respectively. Comparing these
predictions with the most recent experimental value $ M_t = (175.6 \pm
5.5)$ GeV, and recalling that the theoretical values for $M_t$ may
suffer from a correction of less than $\sim 4 \%$ \cite{acta}, we see
that they are consistent with the experimental data.

Turning now to the SSB sector we look for the parameter space in which
the lighter s-tau mass squared $m^2_{\tilde \tau}$ is larger than the
lightest neutralino mass squared $m^2_\chi$ (which is the LSP).  For
the case where all the soft scalar masses are universal at the
unfication scale, there is no region of $M_s=M$ below $O$(few) TeV in
which $m^2_{\tilde \tau} > m^2_\chi$ is satisfied.  But once the
universality condition is relaxed this problem can be solved naturally
(provided the sum rule). More specifically, using the sum rule
(\ref{sumr}) and imposing the conditions a) successful radiative
electroweak symmetry breaking b) $m_{\tilde\tau^2}>0$ and c)
$m_{\tilde\tau^2}> m_{\chi^2}$, we find a comfortable parameter space
for both models (although model $\bf B$ requires large $M\sim 1$ TeV).
The particle spectrum of models $\bf A$ and $\bf B$ in turn is
calculated in terms of 3 and 2 parameters respectively.

In the figure we present  the $m_{\bf 10}$ dependence of 
$m_h$ for for $M= 0.8$ (dashed) $1.0$ (solid) TeV
for the finite Model $\bf B$, which shows that the value of $m_h$ is
stable.  

\begin{figure}
           \epsfxsize= 6 cm   
           \centerline{\epsffile{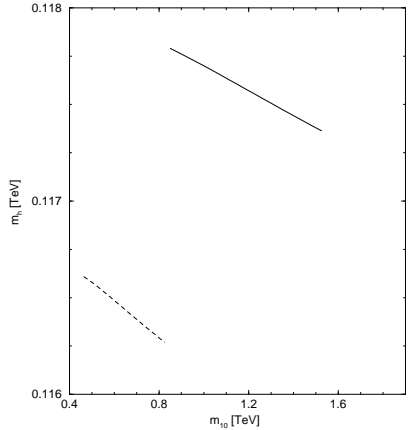}}
           \vspace*{-1cm}
        \caption{ $m_h$ as function of 
$m_{\bf 10}$ for $M= 0.8$ (dashed) $1.0$ (solid) TeV.}
        \label{fig:2}
        \end{figure}

\section{Conclusions}

The search for realistic Finite Unified theories started a few years ago
\cite{kmz1,acta} with the successful prediction of the top quark mass,
and it has now been complemented with a new important ingredient
concerning the finiteness of the SSB sector of the theory. 
Specifically, a sum rule for the soft scalar masses has been obtained
which quarantees the finiteness of the SSB parameters up to two-loops
\cite{kkmz-npb}, avoiding at the same time serious phenomenological
problems related to the previously known ``universal'' solution.  It is
found that this sum rule coincides with that of a certain class of
string models in which the massive string modes are organized into
$N=4$ supermultiplets.  Using the sum rule we can now determine the
spectrum of realistic models in terms of just a few parameters. In
addition to the successful prediction of the top quark mass the
characteristic features of the spectrum are that 1) the lightest Higgs
mass is predicted $\sim 120$ GeV and 2) the s-spectrum starts above
200 GeV. Therefore, the next important test of these Finite Unified
theories will be given with the measurement of the Higgs mass, for
which the models show an appreciable stability in their prediction. In
the figure we show the dependence of the lightest Higgs mass $m_h$ in
terms of the two free parameters of the model $\bf B$.

\end{document}